\documentclass[twocolumn,prl,nobalancelastpage,aps,10pt]{revtex4-1}
\usepackage{graphicx,bm,times}
\usepackage{hyperref}
\usepackage{amsmath}
\hypersetup{
    colorlinks=true,
    linkcolor=blue,
    filecolor=blue,      
    urlcolor=blue,
    citecolor=blue,
    pdftitle={Sharelatex Example},
    pdfpagemode=FullScreen,
    }
\begin{document}

\title{N-body Simulations of the Solar System with CPU-based Parallel Methods}

\author{Tailin Zhu}

\affiliation{School of Physics, University of Bristol}

\begin{abstract}
The gravitational N-body simulation in the Solar system was performed using different parallel approaches with the comparisons in the computational times and speed-up values being carried out under different model sizes and the number of processors. The numerical integration used is a second-order velocity Verlet approach which gives the acceptable accuracy in the orbits of major bodies and asteroids with a time step size of 0.1 days. 
\end{abstract}

\date{February 14, 2020}

\maketitle

\section{1. INTRODUCTION}

N-body simulations can be used to study the dynamics of the orbital system under the Newtonian gravitational forces. A direct approach is using the numerical integration methods to solve the $N$ set of equations of motion, by treating each body as a point mass that interacts with the rest of the $N-1$ bodies \cite{Trenti:2008}. The bodies in the Solar System can be classified as major bodies and small bodies, where the major bodies refer to the Sun, planets, and the major moons, while the small bodies include the asteroids and the minor moons with masses $10^5 - 10^{10}$ times smaller than the Earth. 

For a large model size, parallel programming can significantly reduce the computational time by breaking down the information into smaller pieces and making calculations using multiple processors. A study is carried out in the computation time for simulating the motion of a large number of bodies in the Solar system including the (dwarf-) planets, the moons and a various number of main-belt asteroids. A velocity Verlet integration approach was used to solve the ODEs describing the dynamics of the system. The algorithm was parallelised with two different approaches by accessing the shared memory system and the distributed memory system, where the performance of the two methods will be compared and discussed in detail. 

\section{2. Background}\label{bg}
The equation of motion for a body $p$ under the gravitational forces from all other bodies in a three-dimensional Cartesian coordinate system can be written as
\begin{equation}
    \bm{\ddot{r}}_p = \sum^N_{j=1,j\neq p}Gm_j\frac{\bm{r}_j-\bm{r}_p}{||\bm{r}_j-\bm{r}_p||^3}, \;\;\;p=1,2,...,N,
\end{equation}
where $\bm{r}_p = (x_p,y_p,z_p)$ and $\bm{\ddot{r}}_p$ are the position and acceleration of the $p^{th}$ body, $m_j$ is the mass of the $j^{th}$ body and $G$ is the gravitational constant. This can be solved with given initial conditions, such that \cite{SHARP201689}
\begin{equation}
    \bm{\dot{y}} = f(t,\bm{y}(t)), \;\;\; \bm{y}(t_0) = \bm{y_0}, \;\;\; \Delta t = (t_f-t_0)/N.
\end{equation}
, where $N$ is the number of steps, $t_0$ and $t_f$ are the initial and final time, and $y_0$ is the initial value of the variable. The true solution would have a conserved total energy, and hence can be used to test the accuracy of the numerical solution for a particular time step $\Delta t$ \cite{SHARP201689}. 

There are several popular approaches to solve the ODEs for the interacting classical systems, including Euler, Leapfrog and Verlet \cite{verlet}. The latter two methods are the second-order methods with the error of order $\Delta t^2$, which is better than Euler on first-order only. However, the velocity in the leapfrog does not start with an integer time step makes it difficult to find the initial conditions, since only $\bm{\dot{r}}_0$ is known. Alternatively, it can be expressed with integer time steps to have the same time variables as the positions by using the velocity Verlet method, which is particularly suitable for the energy-conserved system with accelerations depending on positions only \cite{verlet}. There are also higher-order approaches such as $4^{th}$ order Runge-Kutta which can give smaller errors for a certain time step size but requires more computational work to be done and hence longer computation time. With the velocity Verlet method, the set of integration algorithm at time step $i$ can be written as
\begin{align}
    \bm{r}_{i+1,p} &= \bm{r}_{i,p} + \bm{\dot{r}}_{i,p}\Delta t + \frac{1}{2} \bm{\ddot{r}}_{i,p} \Delta t^2, \\
    \bm{\dot{r}}_{i+1,p} &= \bm{\dot{r}}_{i,p} + \frac{1}{2}[\bm{\ddot{r}}_{i,p}+\bm{\ddot{r}}_{i+1,p}]\Delta t.
\end{align}

\begin{figure}[h]
\includegraphics*[width=0.84\linewidth]{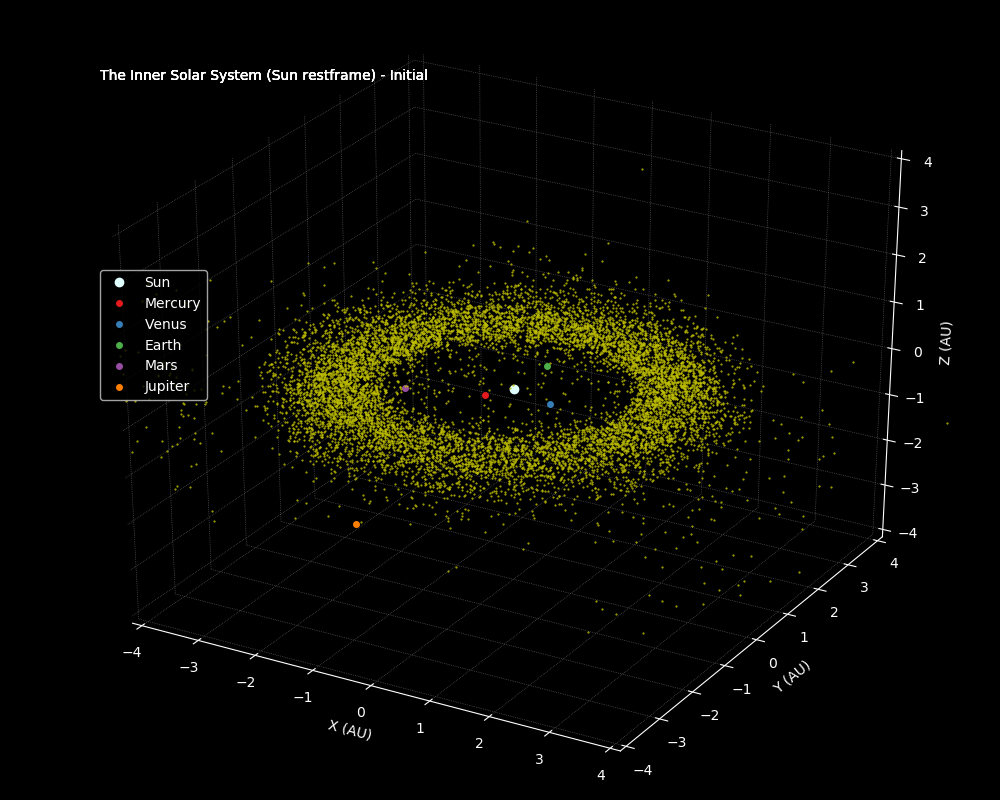}
\caption{The initial position of the Sun, the inner Solar System planets and the main-belt asteroids.}
\label{initial}
\end{figure}

For the structure of the Solar System in this simulation, the main contributed bodies are the Sun and 30 major bodies with masses above $10^{19}$ kg. The extension is to add the main-belt asteroids between the Mars and the Jupiter with a total mass being approximately equal to one-third of the Moon's mass. Only 19 asteroids have the real data in their masses which sums over to give 35$\%$ of the total mass of the main-belt. The rest of the unknown masses are estimated using a uniform distribution between $10^{12}$ kg and $3 \cdot 10^{17}$ kg.
The rest of the objects are filled by over 100 minor moons with masses less than $10^{19}$ kg orbiting around the mars, the pluto and the gas giants. 
The initial positions and velocities for all the major and small bodies were generated from JPL's \texttt{HORIZONS} system on the date 2019-12-22, with the origin being the Sun centre and the $XY$ reference plane being ecliptic and mean equinox of the epoch "J2000". 
Fig.\ref{initial} shows a simulated 3D plot of the initial displacements of 10425 bodies within an area of $\pm 4$AU around the Sun. The yellow dots represent the main-belt asteroids.

\section{3. Parallelisation Methods}\label{pm}

The choice of parallelisation methods depends on the memory system available on a machine. There are two basic types of memory models: shared memory and distributed memory \cite{book}. The shared memory allows all processors to share a global address space which provides uniform and fast data sharing between processors, but the scaling is hard to maintain between memory and processors. The distributed memory assigns each processor an individual address space and hence memory is scalable with processors. However, the communications between the processors should be designed by the programmer.
The supercomputer (\texttt{BlueCrystal Phase3}) used has a distributed memory system with a large number of cluster nodes, with each cluster node containing 16 CPU processors being treated as a shared memory system, and each processor with the associated share of memory being treated as a node. An in-depth investigation on the advantages of the shared memory system with the Open Multi-Processing (OpenMP) and the distributed memory system with the Message Passing Interface (MPI) approach is carried out. With the OpenMP method, the program can be parallelised inside a \texttt{for} loop using a pre-defined algorithm which is easy to programme, but it is limited to run on a single cluster node, and hence only a maximum number of 16 processors can be accessed. With the MPI method, the program can run across different cluster nodes so that hundreds of processors can be assessed. Each processor is manually designed to do calculations for a portion of the bodies, and the results at each time step are gathered together to continue doing calculations for the next time step. For better performance, one processor is taken as the master, and the rest of the processors are the worker. This reduces the number of inputs/outputs by confining them to the master only. The master does its portion of work and sends the instructions to the worker, while the worker receives the instructions from the master and do its portion of work to update the results. 

The MPI approach is suitable for both compiled languages (e.g. C/C++, Fortran) and interpreted languages (e.g. Python), but the OpenMP approach is not supported by Python. Instead of Python, a compiled language called Cython is used to operate OpenMP, where it is a C-like language mostly written in Python by providing data types and calling C functions. 

\section{4. Results \& Discussion}\label{r}

\begin{figure}[h]
\includegraphics*[width=0.96\linewidth]{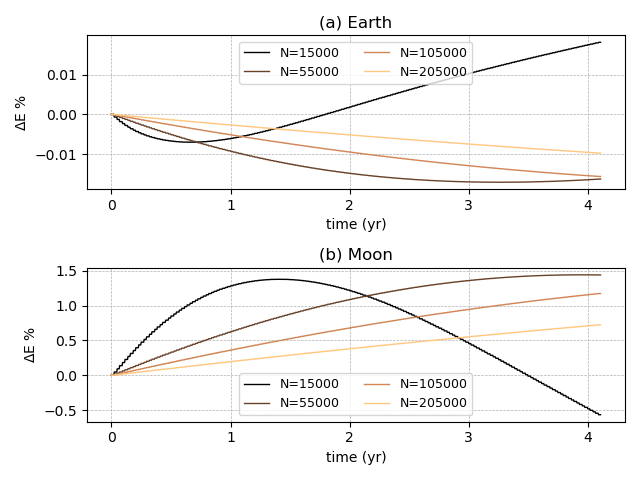}
\caption{The percentage error of the simulations in (a) Earth and (b) Moon orbits for a range of time steps $N$.}
\label{accuracy}
\end{figure}

An accuracy test was done for the results calculated from the velocity Verlet method with a range of time steps. The percentage errors of total energy were simulated over approximately four years (1500 days) with a model size of 5000 bodies. The moons typically need a larger number of time steps than the planets to provide the same accuracy. Here, the percentage errors in the Earth's and the Moon's orbits are presented, as shown in Fig.\ref{accuracy}. The error of the Earth orbit is in between $\pm0.02\%$ for a time step size of about 0.1 days within the four years and gradually converge to smaller errors with the decreasing the time step size, which indicates that the error is acceptable. The error of the Moon orbit is about 70 times larger than the Earth orbit. However, since the Moon has a mass $10^2$ times smaller than the Earth, it interacts less with other bodies under the gravitational force, and hence the influences to the system due to its error become less significant.

\begin{figure}[h]
\includegraphics*[width=1.04\linewidth]{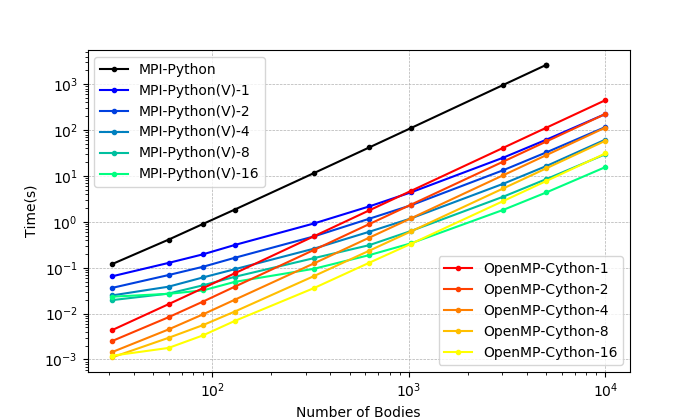}
\caption{The elapsed time for different parallel methods and model sizes. Inside the legend, (V) stands for the vectorised case and the numbers show the number of processors}
\label{body}
\end{figure}

The next part is to study the computation time of the OpenMP and the MPI method for the N-body model in terms of the model sizes and the number of processors using the time step of $N=10$ and the period of one day. Under the MPI method, the simulations were performed using a basic Python script with and without the \texttt{Numpy} vectorisations in calculating the accelerations. The computational time with the non-vectorised MPI approach using one processor is shown in Fig.\ref{body} as a black line, while the vectorised cases are shown as blue lines with a range of processor numbers. By comparing the cases for one processor only, the vectorised MPI code has got an advantage in shortening the time with increased speed and up to about 45 times faster than the non-vectorised case for $10^4$ bodies. The difference is generated by the of usage of \texttt{for} loop in the non-vectorised case, where the \texttt{for} loop takes a long time to check the data types of loop variables, and instead, the vectorised case uses an efficient way to handle the loop internally. The Cythonised code with one thread is always over ten times faster than the non-vectorised Python code with one processor. 

The vectorised MPI approach has a smaller gradient in the increasing of the computational time with the model size compared with the other two approaches. Above a model size of $10^3$, the OpenMP approach exceeds the time taken with the vectorised MPI approach for all the numbers of the processor. Hence, the most efficient way is to use 16 processors with the vectorised MPI approach for a model size greater than $10^3$, and otherwise, use the Cythonised OpenMP method. However, it is worth noting that the computational time is longer for 16 processors at a smaller model size, and additional tests need to be done to choose the exact number of processors.

\begin{figure}[h]
\includegraphics*[width=0.96\linewidth]{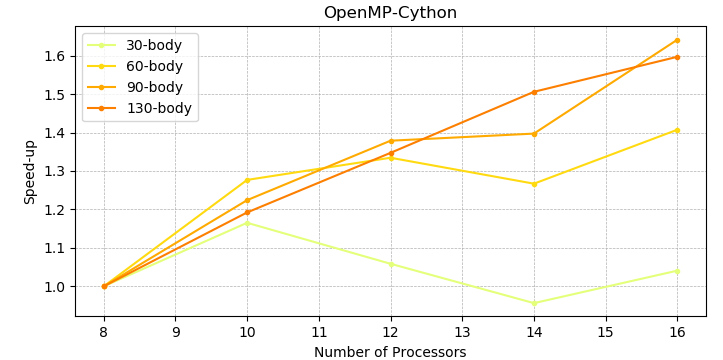}
\caption{The speed-up with different number of processors for small model size with the OpenMP approach.}
\label{cython}
\end{figure}

The relationship between the speed-ups and the number of processors with the OpenMP approach is shown in Fig.\ref{cython} for a small model size, where the speed-up is the ratio of computational time between eight and more than eight processors. Between the processor number 8 and 16,  the maximum speed-up value increases with the model size apart from the 130-body case, which has a different trend as other cases.
It is clear that an over-used number of the processor will not have a significant speed-up and sometimes even increases the computational time. The reason for it is because of the overhead in starting up the new threads, which takes more time than the time reduced in parallelisation.  This would also be due to the effect of Amdahl's law \cite{10.1145/1465482.1465560}, stating that one can only speed the execution up to a limit even if more processors are used.

\begin{figure}[h]
\includegraphics*[width=0.94\linewidth]{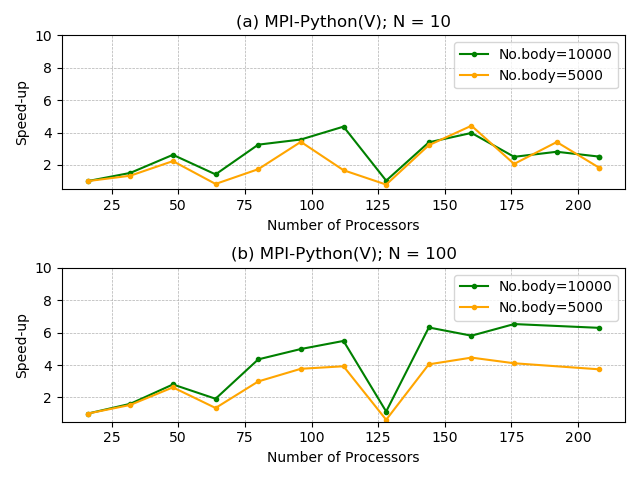}
\caption{The speed-up limit with different number of processors with the vectorised MPI approach.}
\label{processor}
\end{figure}

A further investigation was done on the maximum speed-up for large model sizes with 5000 and $10^4$ bodies using the vectorised MPI approach, as shown in Fig.\ref{processor}. With the time step $N=10$, the computational time oscillates very frequently, which does not show an obvious trend in the speed-up. However, this becomes more obvious for a larger time-step of $10^2$, which takes an overall longer computational time. The speed-up value increases with the number of processors and stops at a processor number of 160 for the 5000-body model and 176 for the $10^4$-body model, which gives a portion of 31 bodies and 56 bodies per processor, respectively. A slightly decreasing in the speed-up can be observed afterwards. This is because of more communications are required for a larger number of processors, and each communication takes time to initiate the transfer of memory, hence cancels the time saved in the parallel calculations. The amount of speed-up increases with the model size and the gap becomes larger as the number of processors increases. It can also be observed that both plots in Fig.\ref{processor} have troughs in the speed-up with 64 and 128 processors, and some even fall below one, meaning that the computational time at these points is longer than the time with one processor. These numbers of processors should be avoided in this simulation. 

\begin{figure}[h]
\includegraphics*[width=0.84\linewidth]{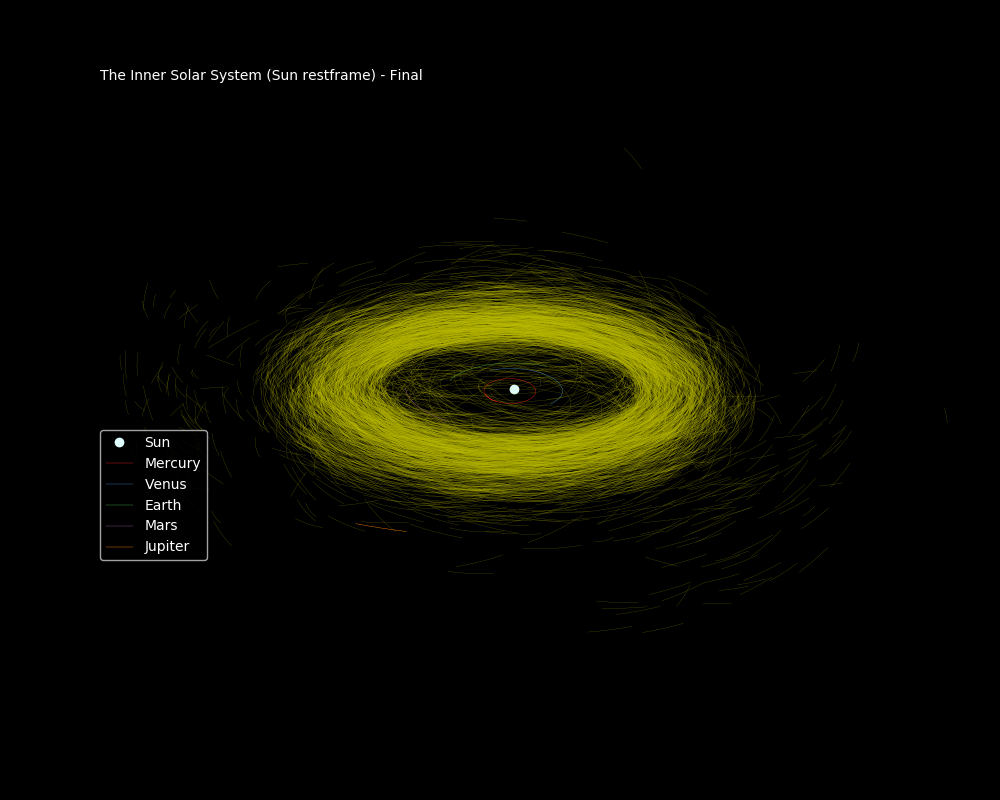}
\caption{The final output for the orbits of the main-belt asteroids and inner Solar planets.}
\label{final}
\end{figure}

The simulation on the dynamics of the Solar System was carried out with $10^4$ bodies for a period of $10^2$ days and the time step of $N = 10^3$ using 176 processors, which takes a computational time of about 23 minutes. This provides relatively correct orbits for the bodies orbiting around the Sun than around the (dwarf-) planets. The animations of the Solar system and planet system dynamics were generated using the \texttt{matplotlib.animation} interface. A final output on the orbits of the bodies in the inner solar system is shown in Fig.\ref{final}, with a distance range of $\pm 4$ AU in the axes.

\section{Conclusions \& Improvements}
In Summary, the choice of the parallel methods and the number of processors are both closely dependent on the model sizes to make an efficient computation. The OpenMP and the vectorised MPI method can reach $10^2$ times faster than the non-vectorised single processor Python approach at the smaller and larger model size, respectively. For the non-vectorised cases, Cython is more efficient than Python. A further improvement can be made using a Cythonised MPI approach. The study on N-body simulation can also be carried out using GPU programming with CUDA threads which could provide a more significant speed-up \cite{SHARP201689}. The simulation can be extended by adding more bodies such as the comets, Kuiper belt asteroids, and Trojan asteroids close to the planets for a longer period.

\def\bibsection{\vskip 6pt \setlength{\parindent}{0pt}{\section{References}} \setlength{\parindent}{12pt}}
\bibliography{references}

\begin{thebibliography}{1}
\expandafter\ifx\csname url\endcsname\relax
  \def\url#1{\texttt{#1}}\fi
\expandafter\ifx\csname urlprefix\endcsname\relax\def\urlprefix{URL }\fi
\providecommand{\bibinfo}[2]{#2}
\providecommand{\eprint}[2][]{\url{#2}}

\bibitem{Trenti:2008}
\bibinfo{author}{Trenti, M.} \& \bibinfo{author}{Hut, P.}
\newblock \bibinfo{title}{{N}-body simulations (gravitational)}.
\newblock \emph{\bibinfo{journal}{Scholarpedia}} \textbf{\bibinfo{volume}{3}},
  \bibinfo{pages}{3930} (\bibinfo{year}{2008}).
\newblock \bibinfo{note}{Revision \#91544}.

\bibitem{SHARP201689}
\bibinfo{author}{Sharp, P.} \& \bibinfo{author}{Newman, W.}
\newblock \bibinfo{title}{{GPU-enabled N-body simulations of the Solar System
  using a VOVS Adams integrator}}.
\newblock \emph{\bibinfo{journal}{J. Comp. Sci.}}
  \textbf{\bibinfo{volume}{16}}, \bibinfo{pages}{89 -- 97}
  (\bibinfo{year}{2016}).

\bibitem{verlet}
\bibinfo{author}{Gould, H.}, \bibinfo{author}{Tobochnik, J.} \&
  \bibinfo{author}{Christian, W.}
\newblock \emph{\bibinfo{title}{{An Introduction to Computer Simulation Methods
  Third Edition (revised)}}} (\bibinfo{year}{2007}).

\bibitem{book}
\bibinfo{author}{Eijkhout, V.}, \bibinfo{author}{van~de Geijn, R.} \&
  \bibinfo{author}{Chow, E.}
\newblock \emph{\bibinfo{title}{{Introduction to High Performance Scientific
  Computing}}} (\bibinfo{year}{2016}).

\bibitem{10.1145/1465482.1465560}
\bibinfo{author}{Amdahl, G.~M.}
\newblock \bibinfo{title}{{Validity of the Single Processor Approach to
  Achieving Large Scale Computing Capabilities}}.
\newblock \bibinfo{pages}{483–485} (\bibinfo{publisher}{Association for
  Computing Machinery}, \bibinfo{address}{New York, NY, USA},
  \bibinfo{year}{1967}).

\end{thebibliography}
\bibliographystyle{naturemag}

\end{document}